Article Thermonuclear corrected 5 18 07

# New AB-Thermonuclear Reactor for Aerospace*

**Alexander Bolonkin**
C&R, 1310 Avenue R, #F-6, Brooklyn, NY 11229, USA
T/F 718-339-4563, aBolonkin@juno.com, http://Bolonkin.narod.ru

## Abstract

About fifty years ago, scientists conducted R&D a thermonuclear reactor that promises a true revolution in the energy industry and, especially, in aerospace. Aircraft can conduct flights of very long distance and for extended periods and that decreases a cost of transportation and allows the saving of ever-more expensive oil-based fuels.

The temperature and pressure required for any particular fuel to fuse is known as the Lawson criterion $L$. Lawson criterion relates to plasma production temperature, plasma density and time. The thermonucler reaction is realised when $L > 10^{14}$.

There are two main methods of nulcear fusion: inertial confinement fusion (ICF) and magnetic confinement fusion (MCF). Existing thermonuclear reactors are very complex, expensive, large, and heavy. They cannot achieve the Lawson creterion.

The author offers an innovation that he suggested early in 1983 for the AB multi-reflex engine and space propulsion (see, for example, a book: A. Bolonkin, Non-Rocket Space Launch and Flight, Elsevier, London, 2006, Chapters 12, 3A). The conventional ICF has a conventional inside surface of its confining shell. This surface absorbs part of the heat radiation from the compressed pellet; the rest of the radiation reflects in all directions and is also absorbed by the shell's interior surface. As result, the target loses energy delivered by outside ignition lasers. This loss is so huge that we need very powerful lasers and we cannot easily or cheaply heat the target to thermonuclear ignition temperature. In all current ICF installations this loss is tens of times more than it is acceptable.

The suggested innovation ICF has on the inside surface of the shell-shaped combustion chamber a covering of small Prism Reflectors (PR) and plasma reflector. These prism reflectors have a noteworthy advantage, in comparison with conventional mirror and especially with conventional shell: they multi-reflect the heat and laser radiation exactly back into collision with the fuel target capsule (pellet). The plasma reflector reflects the Bremsstrahlung radiation.

The offered innovation decreases radiation losses, creates significant radiation pressure and increases the reaction time. The Lawson criterion increases by hundreds of times. The size, cost, and weight of a typical installation will decrease by tens of times.

Currently, the author is researching the efficiency of these innovations.

## Introduction
### Brief information about thermonuclear reactors

**Fusion power** is useful energy generated by nuclear fusion reactions. In this kind of reaction two light atomic nuclei fuse together to form a heavier nucleus and release energy. The largest current experiment, JET, has resulted in fusion power production somewhat larger than the power put into the plasma, maintained for a few seconds. In June 2005, the construction of the experimental reactor ITER, designed to produce several times more fusion power than the power into the plasma over many minutes, was announced. The production of net electrical power from fusion is planned for the next generation experiment after ITER.

Unfortunately, this task is not easy, as scientists thought early. Fusion reactions require a very large amount of energy to initiate in order to overcome the so-called *Coulomb barrier* or *fusion barrier energy*. The key to practical fusion power is to select a fuel that requires the minimum amount of energy to start, that is, the lowest barrier energy. The best fuel from this standpoint is a one-to-one mix of deuterium and tritium; both are heavy isotopes of hydrogen. The D - T (Deuterium & Tritium) mix has a low barrier energy. In order to create the required conditions, the fuel must be heated to tens of millions of degrees, and/or compressed to immense pressures.

At present, D - T is used by two main methods of fusion: inertial confinement fusion (ICF) and magnetic confinement fusion (MCF)(for example, totamak).

In **inertial confinement fusion** (ICF), nuclear fusion reactions are initiated by heating and compressing a target. The target is a pellet that most often contains deuterium and tritium (often only micro or milligrams). Intense laser or ion beams are used for compression. The beams explosively detonate the outer layers of the target. That accelerats the underlying target layers inward, sending a shockwave into the center of pellet mass. If the shockwave is powerful enough and if high enough density at the center is achieved some of the fuel will be heated enough to cause fusion reactions. In a target which has been heated and compressed to the point of thermonuclear ignition, energy can then heat surrounding fuel to cause it to fuse as well, potentially releasing tremendous amounts of energy.

**Magnetic confinement fusion (MCF).** Since plasmas are very good electrical conductors, magnetic fields can also confine fusion fuel. A variety of magnetic configurations can be used, the basic distinction being between magnetic mirror confinement and toroidal confinement, especially tokamaks and stellarators.

**Lawson criterion**. In nuclear fusion research, the Lawson criterion, first derived by John D. Lawson in 1957, is an important general measure of a system that defines the conditions needed for a fusion reactor to reach **ignition**, that is, that the heating of the plasma by the products of the fusion reactions is sufficient to maintain the temperature of the plasma against all losses without external power input. As originally formulated the Lawson criterion gives a minimum required value for the product of the plasma (electron) density $n_e$ and the "energy confinement time" $\tau$. Later analyses suggested that a more useful figure of merit is the "triple product" of density, confinement time, and plasma temperature $T$. The triple product also has a minimum required value, and the name "Lawson criterion" often refers to this inequality.

The key to practical fusion power is to select a fuel that requires the minimum amount of energy to start, that is, the lowest barrier energy. The best fuel from this standpoint is a one-to-one mix of deuterium and tritium; both are heavy isotopes of hydrogen. The D-T (Deuterium & Tritium) mix has a low barrier.

In order to create the required conditions, the fuel must be heated to tens of millions of degrees, and/or compressed to immense pressures. The temperature and pressure required for any particular fuel to fuse is known as the Lawson criterion. For the D - T reaction, the physical value is about

$$L = n_e T \tau > (10^{14} \div 10^{15}) \text{ in "cgs" units or } L = nT\tau > (10^{20} \div 10^{21}) \text{ in CI units},$$

where $T$ is temperature, [KeV], 1 eV = $1.16 \times 10^4$ °K; $n_e$ is matter density, [1/cm$^3$]; $n$ is matter density, [1/m$^3$]; $\tau$ is time, [s]. Last equation is in metric system. The thermonucler rection of $^2$H + $^3$D realises if $L > 10^{20}$ in CI (meter, kilogram, second) units or $L > 10^{14}$ in 'cgs' (cantimetr, gram, second) units.

This number has not yet been achieved in any reactor, although the latest generations of machines have come close. For instance, the reactor TFTR has achieved the densities and energy lifetimes needed to achieve Lawson at the temperatures it can create, but it cannot create those temperatures at the same time. Future ITER aims to do both.

The Lawson criterion applies to inertial confinement fusion as well as to magnetic confinement fusion but is more usefully expressed in a different form. Whereas the energy confinement time in a magnetic system is very difficult to predict or even to establish empirically, in an inertial system it must be on the order of the time it takes sound waves to travel across the plasma:

$$\tau \approx \frac{R}{\sqrt{kT/m_i}}$$

where $\tau$ is time, s; $R$ is distance, m; $k$ is Boltzmann constant; $m_i$ is mass of ion, kg.

Following the above derivation of the limit on $n_e \tau_E$, we see that the product of the density and the radius must be greater than a value related to the minimum of $T^{3/2}/<\sigma v>$ (here $\sigma$ is cross section of reaction, $v$ is ion speed). This condition is traditionally expressed in terms of the mass density $\rho$: $\rho R > 1$ g/cm² .

To satisfy this criterion at the density of solid D + T (0.2 g/cm³) would require an implausibly large laser pulse energy. Assuming the energy required scales with the mass of the fusion plasma ($E_{laser}$ ~ $\rho R^3$ ~ $\rho^{-2}$), compressing the fuel to $10^3$ or $10^4$ times solid density would reduce the energy required by a factor of $10^6$ or $10^8$, bringing it into a realistic range. With a compression by $10^3$, the compressed density will be 200 g/cm³, and the compressed radius can be as small as 0.05 mm. The radius of the fuel before compression would be 0.5 mm. The initial pellet will be perhaps twice as large since most of the mass will be ablated during the compression.

The fusion power density is a good figure of merit to determine the optimum temperature for magnetic confinement, but for inertial confinement the fractional burn-up of the fuel is probably more useful. The burn-up should be proportional to the specific reaction rate ($n^2 <\sigma v>$) times the confinement time (which scales as $T^{1/2}$) divided by the particle density $n$:

$$\text{burn-up fraction} \sim n^2 <\sigma v> T^{1/2} / n \sim (nT) (<\sigma v>/T^{3/2}) .$$

Thus the optimum temperature for inertial confinement fusion is that which maximizes $<\sigma v>/T^{3/2}$, which is slightly higher than the optimum temperature for magnetic confinement.

**Short history of thermonuclear fusion**. One of the earliest (in the late 1970's and early 1980's) serious attempts at an ICF design was **Shiva**, a 20-armed neodymium laser system built at the Lawrence Livermore National Laboratory (LLNL) that started operation in 1978. Shiva was a "proof of concept" design, followed by the **NOVA** design with 10 times the power. Funding for fusion research was severely constrained in the 80's, but NOVA nevertheless successfully gathered enough information for a next generation machine whose goal was ignition. Although net energy can be released even without ignition (the breakeven point), ignition is considered necessary for a *practical* power system.

The resulting design, now known as the National Ignition Facility, commenced being constructed at LLNL in 1997. Originally intended to start construction in the early 1990s, the NIF is now six years behind schedule and overbudget by over $1.4 billion. Nevertheless many of the problems appear to be due to the "big lab" mentality and shifting the focus from pure ICF research to the nuclear stewardship program, LLNLs traditional nuclear weapons-making role. NIF is now scheduled to "burn" in 2010, when the remaining lasers in the 192-beam array are finally installed.

Laser physicists in Europe have put forward plans to build a £500m facility, called HiPER, to study a new approach to laser fusion. A panel of scientists from seven European Union countries believes that a "fast ignition" laser facility could make a significant contribution to fusion research, as well as supporting experiments in other areas of physics. The facility would be designed to achieve high energy gains, providing the critical intermediate step between ignition and a demonstration reactor. It would consist of a long-pulse laser with an energy of 200 kJ to compress the fuel and a short-pulse laser with an energy of 70 kJ to heat it.

Confinement refers to all the conditions necessary to keep a plasma dense and hot long enough to undergo fusion:

- **Equilibrium:** There must be no net forces on any part of the plasma, otherwise it will rapidly disassemble. The exception, of course, is inertial confinement, where the relevant physics must occur faster than the disassembly time.
- **Stability:** The plasma must be so constructed that small deviations are restored to the initial state, otherwise some unavoidable disturbance will occur and grow exponentially until the plasma is destroyed.
- **Transport:** The loss of particles and heat in all channels must be sufficiently slow. The word "confinement" is often used in the restricted sense of "energy confinement".

To produce self-sustaining fusion, the energy released by the reaction (or at least a fraction of it) must be used to heat new reactant nuclei and keep them hot long enough that they also undergo fusion reactions. Retaining the heat generated is called energy **confinement** and may be accomplished in a number of ways.

The hydrogen bomb really has no confinement at all. The fuel is simply allowed to fly apart, but it takes a certain length of time to do this, and during this time fusion can occur. This approach is called **inertial confinement** (fig.1). If more than about a milligram of fuel is used, the explosion would destroy the machine, so controlled thermonuclear fusion using inertial confinement causes tiny pellets of fuel to explode several times a second. To induce the explosion, the pellet must be compressed to about 30 times solid density with energetic beams. If the beams are focused directly on the pellet, it is called **direct drive**, which can in principle be very efficient, but in practice it is difficult to obtain the needed uniformity. An alternative approach is **indirect drive**, in which the beams heat a shell, and the shell radiates x-rays, which then implode the pellet. The beams are commonly laser beams, but heavy and light ion beams and electron beams have all been investigated and tried to one degree or another.

They rely on fuel pellets with a "perfect" shape in order to generate a symmetrical inward shock wave to produce the high-density plasma, and in practice these have proven difficult to produce. A recent development in the field of laser-induced ICF is the use of ultra-short pulse multi-petawatt lasers to heat the plasma of an imploding pellet at exactly the moment of greatest density after it is imploded conventionally using terawatt scale lasers. This research will be carried out on the (currently being built) OMEGA EP petawatt and OMEGA lasers at the University of Rochester and at the GEKKO XII laser at the Institute for Laser Engineering in Osaka Japan which, if fruitful, may have the effect of greatly reducing the cost of a laser fusion-based power source.

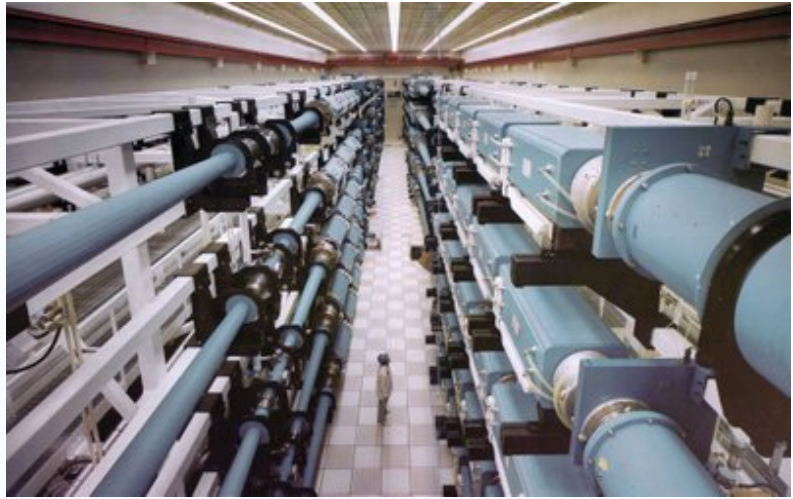

**Fig.1**. Laser installation for inertial thermonuclear reactor.

At the temperatures required for fusion, the fuel is in the form of a plasma with very good electrical conductivity. This opens the possibility to confine the fuel and the energy with magnetic fields, an idea known as **magnetic confinement** (fig.2).

Much of this progress has been achieved with a particular emphasis on tokamaks (fig.2).

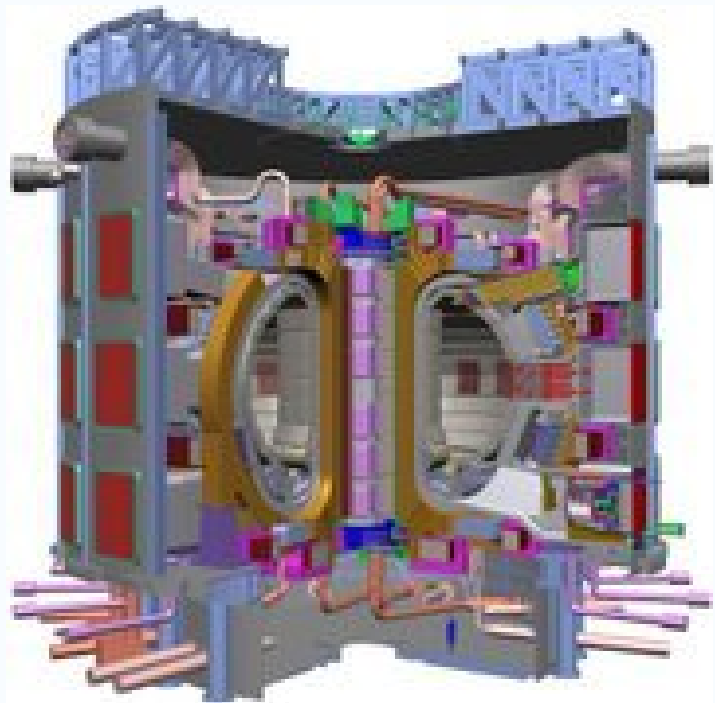

**Fig. 2**. Magnetic thermonuclear reactor. The size of the installation is obvious if you compare it with the "Little Blue Man" inside the machine at the bottom.

In fusion research, achieving a fusion energy gain factor $Q = 1$ is called **breakeven** and is considered a significant although somewhat artificial milestone. **Ignition** refers to an infinite $Q$, that is, a self-sustaining plasma where the losses are made up for by fusion power without any external input. In a practical fusion reactor, some external power will always be required for things like current drive, refueling, profile control, and burn control. A value on the order of $Q = 20$ will be required if the plant is to deliver much more energy than it uses internally.

In a fusion power plant, the nuclear island has a **plasma chamber** with an associated vacuum system, surrounded by a plasma-facing components (first wall and divertor) maintaining the vacuum boundary and absorbing the thermal radiation coming from the plasma, surrounded in turn by a blanket where the neutrons are absorbed to breed tritium and heat a working fluid that transfers the power to the balance of plant. If magnetic confinement is used, a **magnet** system, using primarily cryogenic superconducting magnets, is needed, and usually systems for heating and refueling the plasma and for driving current. In inertial confinement, a **driver** (laser or accelerator) and a focusing system are needed, as well as a means for forming and positioning the **pellets**.

The magnetic fusion energy (MFE) program seeks to establish the conditions to sustain a nuclear fusion reaction in a plasma that is contained by magnetic fields to allow the successful production of fusion power.

In thirty years, scientists have increased the Lawson criterion of the ICF and tokamak installations by tens of times. Unfortunately, all current and some new installations (ICF and totamak) have a Lawrence criterion that is tens of times lower than is necessary (fig.3).

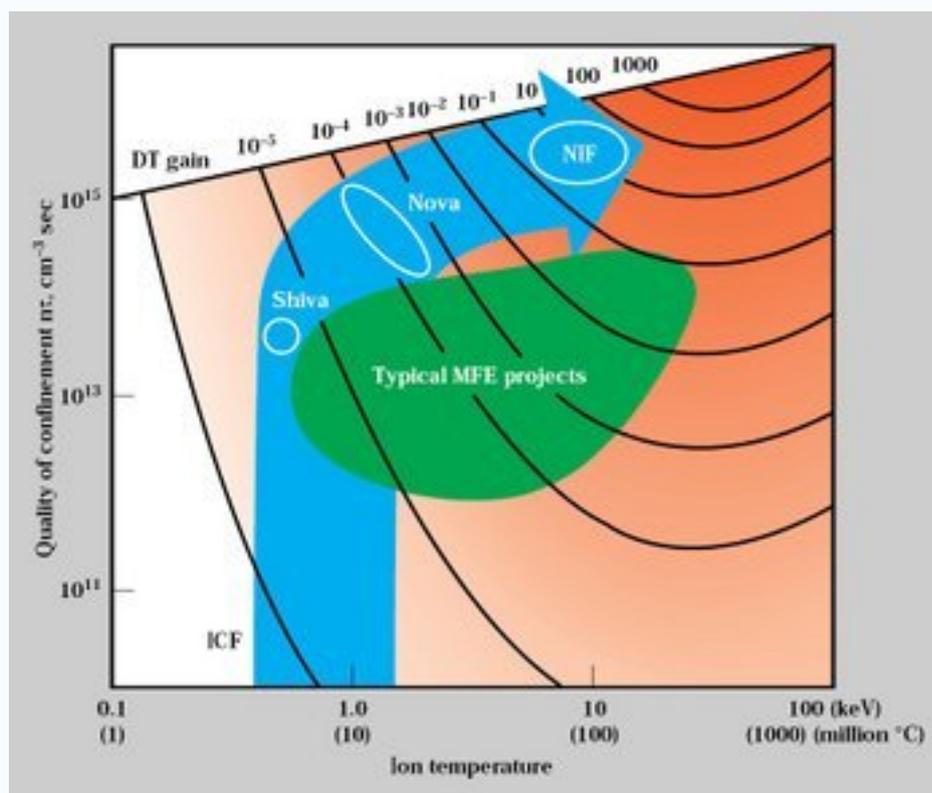

**Fig. 3.** Parameter space occupied by inertial fusion energy and magnetic fusion energy devices. The regime allowing thermonuclear ignition with high gain lies near the upper right corner of the plot.

# Innovation

  As you can see in the equation for thermonuclear reaction (reaction's "ignition") it is necessary to rapidly and greatly increase the temperature, density of target and time to keep the fuel in these induced conditions. In ICF the density of plasma is very high (it increases in 20-30 times in target), the temperature reaches tens of millions °K, but time is measured in nanoseconds. As a result, the Lawson criterion is tens to hundreds of times lower than is required. In a tokomak, the time is parts of second and temperature is tens of millions of degrees, but density of plasma is very small ($10^{20}$ m$^{-3}$). The Lawson criterion is also tens to hundreds of times lower than needed.

  The author offers an innovation suggested early in 1983 [1] and developed later in [2]-[3],[12] for multi-reflex engine and space propulsion. Conventional ICF has conventional inside surface of the shell. This surface absorbs part of the heat radiation emanating from the pellet, the rest of the radiation reflects in all directions and is also absorbed by shell. As result the target loses energy expensively delivered by lasers form outside. This loss is so huge that we need very powerful lasers and we cannot efficiently heat the target to reach ignition temperature (Lawson criterion). In all current ICF installations this loss is tens of times more than is acceptable.

  The innovative ICF has, on its surface inside, a covering of small Prism Reflectors (PR) (fig.4) and plasma layer. The system of prism-plasma reflectors has big advantages in comparison with conventional mirror and especially with conventional shell. The advantages are listed:

  1. The prism-plasma reflector has very high efficiency. The coefficient of its radiate absorption is less about million of times the rate of the conventional mirror for laser wave length. The plasma reflector can reflect the Bremsstrahlung (x-ray) radiation.
  2. The prism mirror reflects the radiation in widely diapason of continuous light spectrum. A conventional mirror reflects the radiation only in narrow diapason of continuous spectrum. That means that any conventional mirror has big absorption of radiation energy even if it has high reflectivity (95%) in narrow interval of the continuous light spectrum.
  3. The prism reflector bounces the laser and heat radiation exactly to a point where heat beam comes up, even if it has defect at position. The conventional mirror having small defect in position (or the pellet is not located exactly in center of sphere) destroys the pellet.
  4. The plasma reflector may be also (and more efficiency) used in cylindrical (toroidal) camera (fig.5) (tokomak, stellarator).

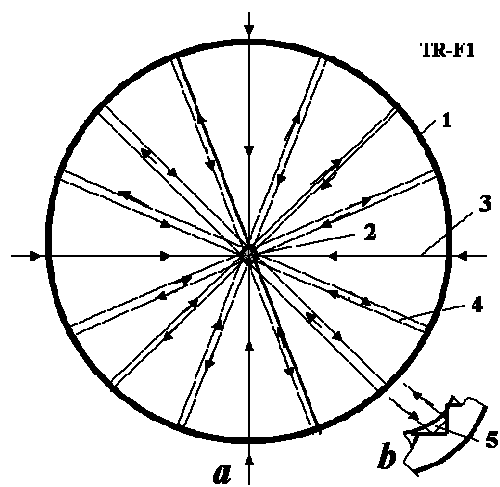

**Fig.4.** Multi-reflex ICF AB thermonuclear reactor. (a) Cross-section of ICF; (b) Cross-section of spherical shell and prism-plasma reflectors. Notations: 1 - target (pellet); 2 - spherical shell; 3- ignition laser beams; 4 - reflected laser-heat beams; 5 - prism-plasma reflector.

The offered Multi-reflex ICF reactor work the following way. The prisms of shell have thin transparency layer. In period when the laser beam is heating the pellet, the beam is multi-reflected from prism and help to heat and to press the pellet. Simultaneously the laser beam heats the prism layer and convert it to plasma. This process is finished to moment when pellet fuel is heated to temperature closed to thermonuclear reaction and begin intensity to produce the Bremsstrahlung radiation from pellet. The Bremsstrahung radiation is multi-reflected by plasma layer of the shell and help to keep the temperature and pressure into pellet.

Look your attention that the reflected plasma of shell can be any (non fuel) plasma and its temperature may be significantly lower temperature of thermonuclear reaction and the density may be other. The aim this plasma is only the reflection of laser and Bremsstrahlung radiation.

The other design of multi-reflection radiation thermonuclear AB reactor is presented in Fig.5. This reactor has cylindrical or toroidal form. In center that contains the gas or solid fuel mix (for example, D + T) into film tube. The internal surface of reactor 1 has internal conductive layer 3. The power electric impulse heats the fuel and simultaneously converts the internal conductive layer 3 into plasma. That is non fuel plasma. Its aim is only multi-reflection of the heat and Bremsstrahlung radiation. Its temperature (and head transferring to reactor walls) is many orders less then fuel plasma.

The multi-refection radiation 5 between central plasma column 2 and cool plasma layer 3 at reactor walls returns the heat to fuel column and help to press the fuel a more time. The electric currency running along the fuel column and into cylindrical plasma reflector creates the circle magnetic field (pinch-effect) which also help to press the fuel plasma and decreases a heat transfer into reactor walls. The additional outer magnet can increase this effect (outer magnet is optional). As reflector can be any (non fuel) plasma and its temperature may be significantly lower temperature of thermonuclear reaction and the density may be higher.

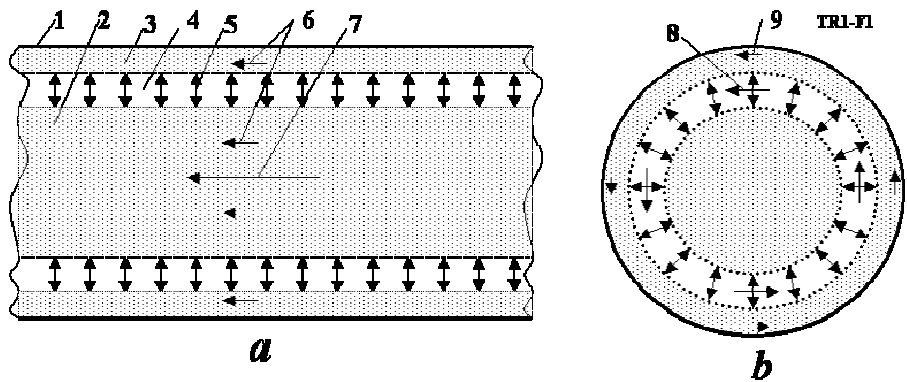

**Fig.5.** Multi-reflection thermonuclear AB reactor. (a) side cross-sectional view; (b) forward cross-sectional view. Notations: 1 - reactor body (cylinder or toroid), 2 - central plasma column (that maybe in thin film), 3 - cylindrical plasma reflector for the central plasma column radiation (it may be separated by thin film) , 4 - vacuum between central plasma column 2 and cylindrical plasma reflector 3 (optinal), 5 - multi-reflection light and Bremsstrahlung radiations, 6 - heating electric currency, 7 - magnetic field from outer magnet (optional), 8 - magnetic field from central plasma column, 9 - magnetic field from the cylindrical plasma tube.

# Theory

## Methods of fuel confinement

**Magnetic confinement**. Magnetic fields can confine fusion fuel because plasma is a very good electrical conductor. A variety of magnetic configurations can be used, the most basic distinction being tokamaks and stellarators.

**Inertial confinement.** A third confinement principle is to apply a rapid pulse of energy to a large part of the surface of a pellet of fusion fuel, causing it to simultaneously "implode" and heat to very high pressure and temperature. If the fuel is dense enough and hot enough, the fusion reaction rate will be high enough to burn a significant fraction of the fuel before it has dissipated. To achieve these extreme conditions, the initially cold fuel must be explosively compressed. Inertial confinement is used in the hydrogen bomb, where the driver is x-rays created by a fission bomb. Inertial confinement is also attempted in "controlled" nuclear fusion, where the driver is a laser, ion, or electron beam.

Some other confinement principles have been investigated, such as muon-catalyzed fusion, the Farnsworth-Hirsch fusor (inertial electrostatic confinement), and bubble fusion.

In man-made fusion, the primary fuel is not constrained to be protons and higher temperatures can be used, so reactions with larger cross-sections are chosen. This implies a lower Lawson criterion, and therefore less startup effort. Another concern is the production of neutrons, which activate the reactor structure radiologically, but also have the advantages of allowing volumetric extraction of the fusion energy and tritium breeding. Reactions that release no neutrons are referred to as *aneutronic*.

In order to be useful as a source of energy, a fusion reaction must satisfy several criteria. It must:

- **be exothermic** - This may be obvious, but it limits the reactants to the low Z (number of protons) side of the curve of binding energy. It also makes helium $^4$He the most common product because of its extraordinarily tight binding, although $^3$He and $^3$H also show up.
- **involve low Z nuclei** - This is because the electrostatic repulsion must be overcome before the nuclei are close enough to fuse.
- **have two reactants** - At anything less than stellar densities, three body collisions are too improbable. It should be noted that in inertial confinement, both stellar densities and temperatures are exceeded to compensate for the shortcomings of the third parameter of the Lawson criterion, ICF's very short confinement time.
- **have two or more products** - This allows simultaneous conservation of energy and momentum without relying on the (weak!) electromagnetic force.
- **conserve both protons and neutrons** - The cross sections for the weak interaction are too small.

Few reactions meet these criteria. The following are those with the largest cross-sections:

**Table 1.** Sutable reactions for thermonuclear fusion.

| | | | | | | | | | | | | |
|---|---|---|---|---|---|---|---|---|---|---|---|---|
| (1) | D | + | T | → | $^4$He (3.5 MeV) | + | n (14.1 MeV) | | | | | |
| (2i) | D | + | D | → | T (1.01 MeV) | + | p (3.02 MeV) | | | | | *50%* |
| (2ii) | | | | → | $^3$He (0.82 MeV) | + | n (2.45 MeV) | | | | | *50%* |
| (3) | D | + | $^3$He | → | $^4$He (3.6 MeV) | + | p (14.7 MeV) | | | | | |
| (4) | T | + | T | → | $^4$He | + 2 | n | + 11.3 MeV | | | | |
| (5) | $^3$He | + | $^3$He | → | $^4$He | + 2 | p | + 12.9 MeV | | | | |
| (6i) | $^3$He | + | T | → | $^4$He | + | p | | + | n | + 12.1 MeV | *51%* |
| (6ii) | | | | → | $^4$He (4.8 MeV) | + | D (9.5 MeV) | | | | | *43%* |
| (6iii) | | | | → | $^4$He (0.5 MeV) | + | n (1.9 MeV) | + | p | (11.9 MeV) | | *6%* |
| (7) | D | + | $^6$Li | → 2 | $^4$He + 22.4 MeV | | | | | | | |

| (8)  | p + $^6$Li → | $^4$He (1.7 MeV) | + | $^3$He (2.3 MeV) |
| (9)  | $^3$He + $^6$Li → 2 $^4$He | | + | p + 16.9 MeV |
| (10) | p + $^{11}$B → 3 $^4$He + 8.7 MeV | | | |

p *(protium),* D *(deuterium), and* T *(tritium) are shorthand notation for the main three isotopes of hydrogen.*

For reactions with two products, the energy is divided between them in inverse proportion to their masses, as shown. In most reactions with three products, the distribution of energy varies. For reactions that can result in more than one set of products, the branching ratios are given.

Some reaction candidates can be eliminated at once. The D - $^6$Li reaction has no advantage compared to p - $^{11}$B because it is roughly as difficult to burn but produces substantially more neutrons through D - D side reactions. There is also a p - $^7$Li reaction, but the cross-section is far too low except possible for $T_i$ > 1 MeV, but at such high temperatures an endothermic, direct neutron-producing reaction also becomes very significant. Finally there is also a p - $^9$Be reaction, which is not only difficult to burn, but $^9$Be can be easily induced to split into two alphas and a neutron.

In addition to the fusion reactions, the following reactions with neutrons are important in order to "breed" tritium in "dry" fusion bombs and some proposed fusion reactors:

$$n + {}^6Li \rightarrow T(2.7 \text{ MeV}) + {}^4He(2.1 \text{ MeV}),$$
$$n + {}^7Li \rightarrow T + {}^4He + n\,.$$

To evaluate the usefulness of these reactions, in addition to the reactants, the products, and the energy released, one needs to know something about the cross section. Any given fusion device will have a maximum plasma pressure that it can sustain, and an economical device will always operate near this maximum. Given this pressure, the largest fusion output is obtained when the temperature is selected so that <σv>/T² is a maximum. This is also the temperature at which the value of the triple product $nT\tau$ required for ignition is a minimum. This chosen optimum temperature and the value of <σv>/T² at that temperature is given for a few of these reactions in the following table.

**Table 2**. Optimum temperature and the value of <σv>/T² at that temperature

| fuel | *T* [keV] | <σv>/T² [m³/s/keV²] |
|---|---|---|
| D-T | 13.6 | 1.24×10$^{-24}$ |
| D-D | 15 | 1.28×10$^{-26}$ |
| D-$^3$He | 58 | 2.24×10$^{-26}$ |
| p-$^6$Li | 66 | 1.46×10$^{-27}$ |
| p-$^{11}$B | 123 | 3.01×10$^{-27}$ |

Note that many of the reactions form chains. For instance, a reactor fueled with T and $^3$He will create some D, which is then possible to use in the D + $^3$He reaction if the energies are "right". An elegant idea is to combine the reactions (8) and (9). The $^3$He from reaction (8) can react with $^6$Li in reaction (9) before completely thermalizing. This produces an energetic proton which in turn undergoes reaction

(8) before thermalizing. A detailed analysis shows that this idea will not really work well, but it is a good example of a case where the usual assumption of a Maxwellian plasma is not appropriate.

Any of the reactions above can, in principle, be the basis of fusion power production. In addition to the temperature and cross section discussed above, we must consider the total energy of the fusion products $E_{fus}$, the energy of the charged fusion products $E_{ch}$, and the atomic number $Z$ of the non-hydrogenic reactant.

Specification of the D - D reaction entails some difficulties, though. To begin with, one must average over the two branches (2) and (3). More difficult is to decide how to treat the T and $^3$He products. T burns so well in a deuterium plasma that it is almost impossible to extract from the plasma. The D - $^3$He reaction is optimized at a much higher temperature, so the burn-up at the optimum D - D temperature may be low, so it seems reasonable to assume the T but not the $^3$He gets burned up and adds its energy to the net reaction. Thus we will count the D - D fusion energy as $E_{fus}$ = (4.03+17.6+3.27)/2 = 12.5 MeV and the energy in charged particles as $E_{ch}$ = (4.03+3.5+0.82)/2 = 4.2 MeV.

Another unique aspect of the D - D reaction is that there is only one reactant, which must be taken into account when calculating the reaction rate.

With this choice, we tabulate parameters for four of the most important reactions.

**Table 3.** Parameters of the most important reactions.

| fuel | Z | $E_{fus}$ [MeV] | $E_{ch}$ [MeV] | neutronicity |
|---|---|---|---|---|
| D-T | 1 | 17.6 | 3.5 | 0.80 |
| D-D | 1 | 12.5 | 4.2 | 0.66 |
| D-$^3$He | 2 | 18.3 | 18.3 | ~0.05 |
| p-$^{11}$B | 5 | 8.7 | 8.7 | ~0.001 |

The last column is the **neutronicity** of the reaction, the fraction of the fusion energy released as neutrons. This is an important indicator of the magnitude of the problems associated with neutrons like radiation damage, biological shielding, remote handling, and safety. For the first two reactions it is calculated as ($E_{fus}$ - $E_{ch}$)/$E_{fus}$. For the last two reactions, where this calculation would give zero, the values quoted are rough estimates based on side reactions that produce neutrons in a plasma in thermal equilibrium.

Of course, the reactants should also be mixed in the optimal proportions. This is the case when each reactant ion plus its associated electrons accounts for half the pressure. Assuming that the total pressure is fixed, this means that density of the non-hydrogenic ion is smaller than that of the hydrogenic ion by a factor 2/(Z+1). Therefore the rate for these reactions is reduced by the same factor, on top of any differences in the values of <σv>/T². On the other hand, because the D - D reaction has only one reactant, the rate is twice as high as if the fuel were divided between two hydrogenic species.

Thus, there is a "penalty" of (2/(Z+1)) for non-hydrogenic fuels arising from the fact that they require more electrons, which take up pressure without participating in the fusion reaction. There is, at the same time, a "bonus" of a factor 2 for D - D due to the fact that each ion can react with any of the other ions, not just a fraction of them.

We can now compare these reactions in the following table 4.

Table 4. Comparison of reactions

| fuel | $\langle\sigma v\rangle/T^2$ | penalty/bonus | reactivity | Lawson criterion | power density |
|---|---|---|---|---|---|
| D-T | $1.24\times10^{-24}$ | 1 | 1 | 1 | 1 |
| D-D | $1.28\times10^{-26}$ | 2 | 48 | 30 | 68 |
| D-$^3$He | $2.24\times10^{-26}$ | 2/3 | 83 | 16 | 80 |
| p-$^{11}$B | $3.01\times10^{-27}$ | 1/3 | 1240 | 500 | 2500 |

The maximum value of $\langle\sigma v\rangle/T^2$ is taken from a previous table. The "penalty/bonus" factor is that related to a non-hydrogenic reactant or a single-species reaction. The values in the column "reactivity" are found by dividing ($1.24\times10^{-24}$ by the product of the second and third columns. It indicates the factor by which the other reactions occur more slowly than the D-T reaction under comparable conditions. The column "Lawson criterion" weights these results with $E_{ch}$ and gives an indication of how much more difficult it is to achieve ignition with these reactions, relative to the difficulty for the D-T reaction. The last column is labeled "power density" and weights the practical reactivity with $E_{fus}$. It indicates how much lower the fusion power density of the other reactions is compared to the D-T reaction and can be considered a measure of the economic potential.

## Bremsstrahlung losses

Bremsstrahlung, (from the German *bremsen*, to brake and *Strahlung*, radiation, thus, "braking radiation"), is electromagnetic radiation produced by the acceleration of a charged particle, such as an electron, when deflected by another charged particle, such as an atomic nucleus. The term is also used to refer to the process of producing the radiation. Bremsstrahlung has a continuous spectrum. The phenomenon was discovered by Nikola Tesla (1856-1943) during high frequency research he conducted between 1888 and 1897.

Bremsstrahlung may also be referred to as free-free radiation. This refers to the radiation that arises as a result of a charged particle that is free both before and after the deflection (acceleration) that causes the emission. Strictly speaking, bremsstrahlung refers to any radiation due to the acceleration of a charged particle, which includes synchrotron radiation; however, it is frequently used (even when not speaking German) in the more literal and narrow sense of radiation from electrons stopping in matter.

The ions undergoing fusion will essentially never occur alone but will be mixed with electrons that neutralize the ions' electrical charge and form a plasma. The electrons will generally have a temperature comparable to or greater than that of the ions, so they will collide with the ions and emit Bremsstrahlung. The Sun and stars are opaque to Bremsstrahlung, but essentially any terrestrial fusion reactor will be optically thin at relevant wavelengths. Bremsstrahlung is also difficult to reflect and difficult to convert directly to electricity, so the ratio of fusion power produced to Bremsstrahlung radiation lost is an important figure of merit. This ratio is generally maximized at a much higher temperature than that which maximizes the power density (see the previous subsection). The following table shows the rough optimum temperature and the power ratio at that temperature for several reactions.

Table 5. Rough optimum temperature and the power ratio of fusion and Bremsstrahlung radiation lost.

| fuel | $T_i$ (keV) | $P_{fusion}/P_{Bremsstrahlung}$ |
|---|---|---|
| D-T | 50 | 140 |
| D-D | 500 | 2.9 |
| D-$^3$He | 100 | 5.3 |
| $^3$He-$^3$He | 1000 | 0.72 |
| p-$^6$Li | 800 | 0.21 |
| p-$^{11}$B | 300 | 0.57 |

The actual ratios of fusion to Bremsstrahlung power will likely be significantly lower for several reasons. For one, the calculation assumes that the energy of the fusion products is transmitted completely to the fuel ions, which then lose energy to the electrons by collisions, which in turn lose energy by Bremsstrahlung. However because the fusion products move much faster than the fuel ions, they will give up a significant fraction of their energy directly to the electrons. Secondly, the plasma is assumed to be composed purely of fuel ions. In practice, there will be a significant proportion of impurity ions, which will lower the ratio. In particular, the fusion products themselves *must* remain in the plasma until they have given up their energy, and *will* remain some time after that in any proposed confinement scheme. Finally, all channels of energy loss other than Bremsstrahlung have been neglected. The last two factors are related. On theoretical and experimental grounds, particle and energy confinement seem to be closely related. In a confinement scheme that does a good job of retaining energy, fusion products will build up. If the fusion products are efficiently ejected, then energy confinement will be poor, too.

The temperatures maximizing the fusion power compared to the Bremsstrahlung are in every case higher than the temperature that maximizes the power density and minimizes the required value of the fusion triple product (Lawson criterion). This will not change the optimum operating point for D-T very much because the Bremsstrahlung fraction is low, but it will push the other fuels into regimes where the power density relative to D-T is even lower and the required confinement even more difficult to achieve. For D-D and D-$^3$He, Bremsstrahlung losses will be a serious, possibly prohibitive problem. For $^3$He-$^3$He, p-$^6$Li and p-$^{11}$B the Bremsstrahlung losses appear to make a fusion reactor using these fuels impossible.

In a plasma, the free electrons are constantly producing Bremsstrahlung in collisions with the ions. The power density of the Bremsstrahlung radiated is given by

$$P_{Br} = \frac{16\alpha^3 h^2}{\sqrt{3}\, m_e^{3/2}} n_e^2 T_e^{1/2} Z_{eff},$$

$T_e$ is the electron temperature, $\alpha$ is the fine structure constant, $h$ is Planck's constant, and the "effective" ion charge state $Z_{eff}$ is given by an average over the charge states of the ions:

$$Z_{eff} = \Sigma (Z^2 n_Z) \,.\, n_y \text{ ю}$$

This formula is derived in "Basic Principles of Plasmas Physics: A Statistical Approach" by S. Ichimaru, p. 228. It applies for high enough $T_e$ that the electron deBroglie wavelength is longer than the classical Coulomb distance of closest approach. In practical units, this formula gives

$$P_{Br} = (1.69 \times 10^{-32} \text{ W cm}^{-3}) (n_e/\text{cm}^{-3})^2 (T_e/\text{eV})^{1/2} Z_{eff}$$
$$= (5.34 \times 10^{-37} \text{ W m}^{-3}) (n_e/\text{m}^{-3})^2 (T_e/\text{keV})^{1/2} Z_{eff}$$

For very high temperatures there are relativistic corrections to this formula, that is, additional terms of order $T_e/m_e c^2$.

# List of main equations.

Below are the main equations for estimation of benefits from the offered innovation.

1. **Energy needed for ignition** and inserted in inertial fusion reactor must be more than

$$W > \frac{10^8}{\eta^3 \alpha^2}, \qquad (1)$$

where $W$ is energy, J; $\eta$ is coefficient of efficiency of reaction, $\alpha$ is compression ratio of target (currently, it is 10 - 30).

Fig. 6 shows a magnitude $n\tau$ (analog of Lawson criterion) required for ignition.

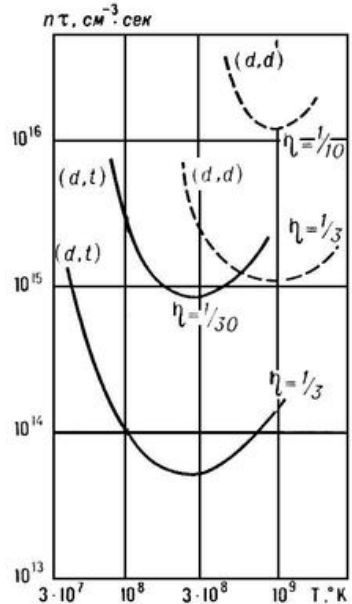

**Fig. 6.** Analog of Lawson criterion for thermonuclear reaction. Value $n\tau$ is in s/cm$^3$.

At present the industry produces power lasers:
- Carbon dioxide lasers emit up to 100 kW at 9.6 μm and 10.6 μm, and are used in industry for cutting and welding.
- Carbon monoxide lasers must be cooled but can produce up to 500 kW.

Special laser:

- As of 2005 the National Ignition Facility is working on a system that, when complete, will contain

    192-beam, 1.8-megajoule, 700-terawatt laser system adjoining a 10-meter-diameter target chamber.

- 1.25 PW - world's most powerful **laser** (claimed on 23 May 1996 by Lawrence Livermore Laboratory).

**2. Radiation energy from hot solid black body** is (Stefan-Boltzmann Law):

$$E = \sigma T^4, \qquad (2)$$

where $E$ is emitted energy, W/m$^2$; $\sigma = 5.67 \times 10^{-8}$ - Stefan-Boltzmann constant, W/m$^2$ °K$^4$; $T$ is temperature in °K.

**3. Wave length** corresponded of maximum energy density (Wien's Law) is

$$\lambda_0 = \frac{b}{T}, \quad \omega = \frac{2\pi}{\lambda_0}, \qquad (3)$$

where $b = 2.8978 \times 10^{-3}$ is constant, m °K; $T$ is temperature, °K; $\omega$ is angle frequency of wave, rad/s.

**4. Pressure for full reflection** is

$$F = 2E/c, \qquad (3)$$

where $F$ - pressure, N/m$^2$; $c = 3 \times 10^8$ is light, m/s, $E$ is radiation power, W/m$^2$. If matter does not reflect, the pressure equals

$$F = E/c.$$

**5. Pressure for multi-reflection** [2-3] is

$$F = \frac{2E}{c}\left(\frac{1}{1-q}\right), \qquad (4)$$

where $q$ is reflection coefficient. For example, if $q = 0.98$ the radiation pressure increases in 50 times.

**6. The Bremsstrahlung loss** energy of plasma by radiation is ($T > 10^6$ °K) is

$$P_{Br} = 5.34 \cdot 10^{-37} n_e^2 T^{0.5} Z_{eff}, \quad \text{where} \quad Z_{eff} = \sum (Z^2 n_z)/n_e, \qquad (5)$$

where $P_{Br}$ is power of Bramsstrahlung radiation, W/m$^2$; $n_e$ is number of particles in m$^3$; $T$ is a plasma temperature, KeV; $Z$ is charge state; $Z_{eff}$ is cross-section coefficient for multi-charges ions. For reaction H+H, H+D, D+T the $Z_{eff}$ closes 1.

That loss may be very much. For some reaction they are more then useful nuclear energy and fusion nuclear reaction may be stopped. The Bremsstrahling emission has continuous spectra.

Unfortunately, the author cannot find data about Bremsstrahling spectra. The energy of Bremsstrahlung photon equals energy of plazma electron. The formula for probability wave length (5a) below is received by author:

From $E = kT_k = h\nu$, $\lambda = \dfrac{c}{\nu}$ we receive $\lambda_m = \dfrac{ch}{kT_k} = \dfrac{0.0144}{T_k}, \quad \omega = \dfrac{2\pi}{\lambda_m} = 436 T_k,$ (5a)

where $E$ is electron energy, J; $k = 1{,}38 \times 10^{-23}$ is Boltzmann's constant, J/°K; $h = 6.525 \times 10^{-34}$ is Plank's constant, J·s ; $c = 3 \times 10^8$ is light speed, m/s; $T_k$ is temperature in °K; $\lambda_m$ is wave length, m; $\nu$ is wave frequency, 1/s; $\omega$ is wave frequency in rad/s.

For example, for $T_k = 10^8$ °K the $\lambda_m = 1.44 \times 10^{-10}$ m. That value is in x-ray diapason $\lambda > 10^{-10}$ m. For very high temperature a part of this spectrum may be in the soft x-ray region, but x-ray is difficult reflected and retracted by conventional methods.

**7. Electron frequency in plasma** is

$$\omega_0 = \left(\frac{4\pi n_e e^2}{m_e}\right)^{1/2}, \quad \text{or} \quad \omega_0 = 5.64 \times 10^4 (n_e)^{1/2} \quad \text{in "cgs" units, or} \quad \omega_0 = 56.4(n)^{1/2} \quad \text{in CI units} \tag{6}$$

where $\omega_0 = \omega_{pe}$ is electron frequency, rad/s; $n_e$ is electron density, [1/cm$^3$]; $n$ is electron density, [1/m$^3$]; $m_e = 9.11 \times 10^{-28}$ is mass of electron, g; $e = 1.6 \times 10^{-19}$ is electron charge, C.

The plasma is reflected an electromagnet radiation if frequency of electromagnet radiation is less then electron frequency in plasma, $\omega < \omega_0$. That reflectivity is high for $T > 15 \times 10^6$ °K it is more then silver and increases with plasma temperature as $T^{3/2}$. The frequency in Hz is $f_{pe} = \omega_{pe}/2\pi$.

**8. The deep of penetration** of outer radiation into plasma is

$$d_p = \frac{c}{\omega_{pe}} = 5.31 \cdot 10^5 n_e^{-1/2} . \tag{7}$$

**9. The gas (plasma) dynamic pressure**, $p$, is

$$p = nk(T_e + T_i). \quad \text{if} \quad T_e = T_i = T, \quad \text{then} \quad p = 2nkT , \tag{8}$$

**10. The gas (plasma) ion pressure**, $p$, is

$$p = \frac{2}{3} nkT , \tag{9}$$

where $k = 1.38 \times 10^{-23}$ is Boltzmann constant; $T_e$ is temperature of electrons, °K; $T_i$ is temperature of ions, °K. These temperatures may be different. The $n$ is plasma density in 1/m$^3$.

**11. The magnetic and electrostatic pressure**, $p$, are

$$p = \frac{B^2}{2\mu_0}, \quad p = \frac{1}{2}\varepsilon_0 E_S^2 , \tag{10}$$

where $B$ is electromagnetic induction, Tesla; $\mu_0 = 4\pi \times 10^{-7}$ electromagnetic constant; $\varepsilon_0 = 8.85 \times 10^{-12}$, F/m, is electrostatic constant; $E_S$ is electrostatic intensity, V/m.

**12. Ion thermal velocity** is

$$v_{Ti} = \left(\frac{kT_i}{m_i}\right)^{1/2} = 9.79 \times 10^5 \mu^{-1/2} T_i^{1/2} \quad \text{cm/s} \tag{11}$$

where $\mu = m_i/m_p$, $m_i$ is mass of ion, kg; $m_p = 1.67 \times 10^{-27}$ is mass of proton, kg.

## Computation (estimation) of innovation

The offered innovation is a special prism, a high reflectivity mirror that returns the laser and heat radiation exactly to its point of origination. The other innovation is special not hot plasma layer which reflects the Bremsstrahlung (x-ray) radiation. As a result, the energy returns in fuel plasma, the plasma radiation presses the plasma and impedes or does not allow its expansion. The plasma can have high reflectivity and this effect may be increased by tens and hundreds times.

For estimation possibility of innovation we compute the radiation pressure, the condition of plasma reflection and compare them with dynamic pressure of plasma.

1. **Fuel density**. The particles (ions) density of hydrogen in 1 m$^3$ equals

$$n = M/m_i ,$$

where $M$ is fuel mass density, kg/m$^3$; $m_i = 2 \times 1.672 \times 10^{-27}$ kg for hydrogen. The $n \approx 10^{20}$ 1/m$^3$ in the magnetic confinement fusion reactor; $n \approx 5.2 \times 10^{25}$ 1/m$^3$ ($M = 0.0875$ kg/m$^3$ for gas hydrogen in a pressure 1 atm, $T = 288$ °K (for other pressure the $n$ must be changed in same times)); $n \approx 4.25 \times 10^{28}$ $1/m^3$  ($M = 71$  kg/m$^3$ for liquid hydrogen in a pressure 1 atm, (in conventional inertial confinement fusion reactor, the hydrogen density may be more in 10 - 30 times, under a conventional reactive pressure of a fuel capsule cover)).

1. **Electron plasma frequency**. Electron frequency of plasma is computed by equation (6). For $n \approx 10^{20}$ $1/m^3$ that is equals $\omega_0 = 5.64 \times 10^{11}$ rad/s, for $n \approx 10^{28}$ $1/m^3$  that is equals $\omega_0 = 5.64 \times 10^{15}$ rad/s .
2. **Plasma skin depth**, the depth in a plasma to which an electromagnetic radiation can penetrate (Eq. (7)) is: For $n \approx 10^{20}$ $1/m^3$ that is equals $d_p = 5.31 \times 10^{-2}$ cm, for $n \approx 10^{28}$ $1/m^3$  that is equals $d_p = 5.31 \times 10^{-6}$ cm. As you see the depth is small.
3. **Estimation of maximum frequency** of Bremsstrahlung radiation. The Bremsstrahlung radiation has a continuous spectrum. The author cannot find data about it. He offers the spectrum has the Planck's law. That means about 80% of radiation energy has low frequency over at maximum. For primary estimation we use the eq. (5a). For $T \approx 10^6$ °K that is equals $\omega = 4.36 \times 10^8$ rad/s, For $T \approx 10^8$ °K that is equals $\omega = 4.36 \times 10^{10}$ rad/s. It is less then minimum electron frequency for very rarely ($n = 10^{20}$ $1/m^3$) plasma $\omega_0 = 5.64 \times 10^{11}$ rad/s [eq.(6)}. That means the plasma will reflect the Bremsstrahlung radiation. The minimum of Bremsstrahlung length wives ($\lambda_0 = 1.44 \times 10^{-10}$ m, $T \approx 10^8$ °K) locates in soft x-ray region (x-ray region is $\lambda_0 = (0,06 \div 120) \times 10^{-10}$ m).  They can be reflected by the offered plasma reflector.
4. **Coefficient reflectivity of plasma**. No data about plasma reflectivity. However, from general theory of reflectivity it is known the reflectivity depends from conductivity. Silver has the best conductivity from solid body and best reflectivity. It is about $q = 0.78 \div 0.99$ (it depends from frequency of radiation). The plasma for $T > 15 \times 10^6$ °K has better conductivity then silver. The plasma conductivity increases as $T^{3/2}$. That means the plasma having the $T \approx 10^8$ °K has reflectivity in 17.2 times better then silver. The efficiency of offered innovation very strong depends from reflectivity of plasma. The reflectivity of prism mirror in laser beam diapason is very high [2]. We neglect the loss in it.
5. **Bremsstrahlung radiation**. These computations are presented in Fig. 7 below.
6. 

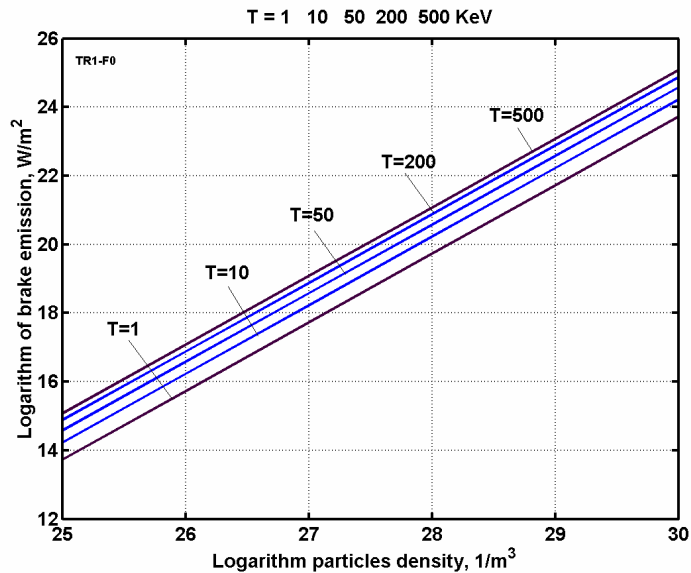

**Fig. 7**. Logarithm of brake (Bremsstrahlung) radiation via particles density.

7. **The radiation and opposed gas kinetic pressure** of plasma is presented in Figs. 8 - 10.

Fig. 8 shows if a plasma reflectivity is zero (*q* = 0, that means the laser and Bremsstrahlung radiation are fully adsorbed by fuel plasma after first reflection by prism and plasma reflectors), the Bremsstrahlung pressure in offered reactor is more the a gas kinetic pressure when a plasma density logarithm is more 29.

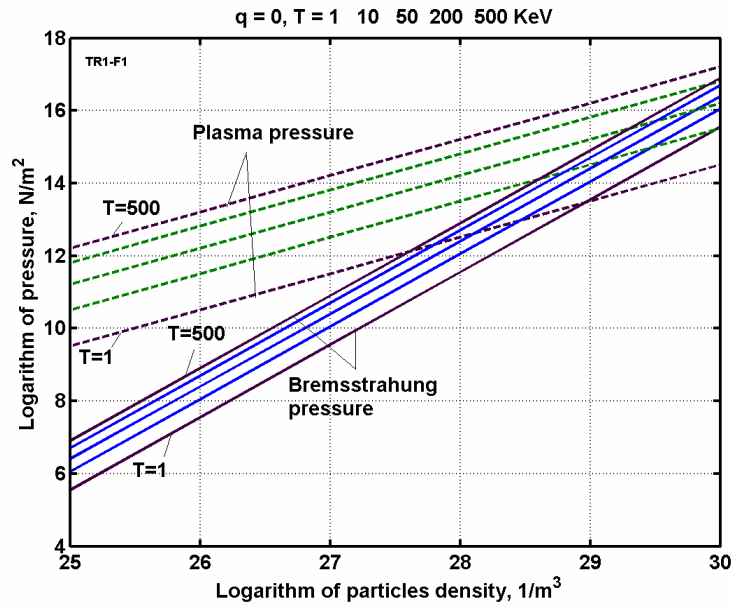

**Fig. 8**. Logarithm of brake (Bremsstrahlung) and plasma pressure via particles density for zero fuel plasma reflectivity
(*q* = 0. Fuel plasma is black body).

Fig.9 at below is computed for reflectivity *q* = 0.9 (ten of full reflections). The Bremsstrahlung pressure in offered reactor is more gas kinetic pressure when the plasma density logarithm is more 28.

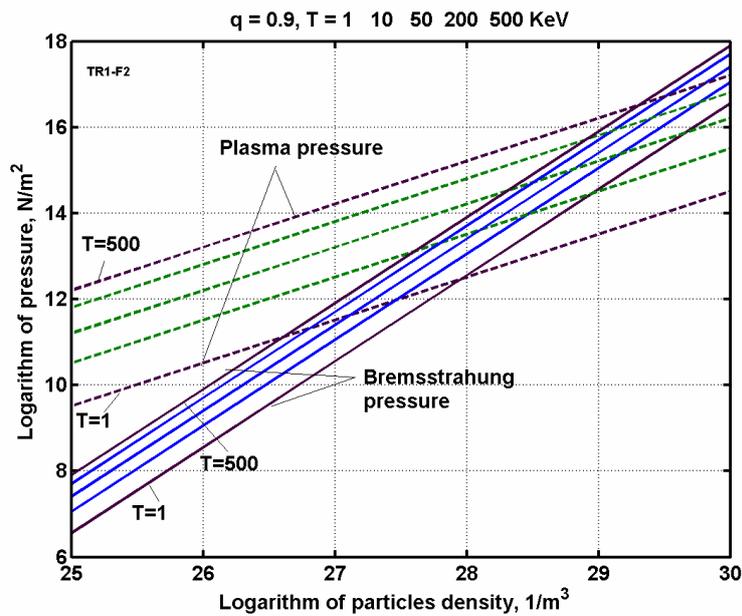

**Fig. 9**. Logarithm of brake (Bremsstrahlung) and particles pressure via particles density for plasma reflectivity
*q* = 0.9.

Fig.10 below is computed for reflectivity $q = 0.99$ (hundred of full reflections). The Bremsstrahlung pressure in offered reactor is more the gas kinetic pressure when the fuel plasma density logarithm is more 27.

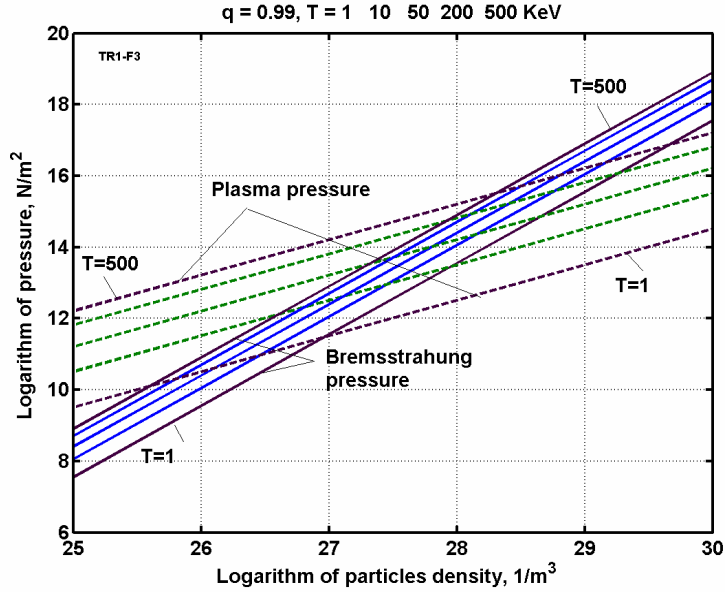

**Fig. 10**. Logarithm of brake (Bremsstrahlung) and particles pressure via particles density for plasma reflectivity $q = 0.99$.

Notice that the Bremsstrahlung pressure is very high. For example, if $n = 10^{29}$ 1/m$^3$ the radiation pressure equals $10^{16}$ N/m$^2$ = $10^{11}$ atmospheres (100 billions atm!). For fuel density more $n = 10^{27}$ the radiation pressure may be 2 - 3 order more then a gas dynamic plasma pressure.

**8. Some other estimates**.

*The energy needs for heating* of a mixture D + T for low temperature $T < 2 \times 10^3$ °K and coefficient of efficiency = 1 approximately equals $E \approx MC_vT = 10^5$ J, where $C_v \approx 10^3$ J/kg is average thermal capacity of fuel, $M$ is mass in kg. For high temperature $E = knT$. That $E$ is small, but a thermal loss is very high.

*The drive laser beam pressure*. If drive laser beam has power $W = 1$ MW and a surface of the fuel capsule (pellet) equals $S = 5$ mm$^2$, the laser beam pressure (in full reflection) $P_L = 2W/(cS) = 6.7 \times 10^2$ N/m$^2$. The double prism mirror (in inside of combustion chamber and on the cover of fuel capsule) can increase it in $10 \div 10^6$ times. But ii is less then gas dynamic pressure of plasma in many times. For W= 1 GW, the laser pressure is $P_L = 6.7 \times 10^5$ N/m$^2$. That may be increased many times by the prism reflector.

*The reactive pressure*. From Eq.(11) we can estimate that the ion speed for $T = 10^8$ °K. That is approximately $V = 900$ km/s. If $M = 0.1$ $\mu$g $= 10^{-7}$ kg of a mixture D+T is increased their speed to this value in time $\tau = 10^{-9}$ s, the reactive force will be $F = MV/\tau \approx 10^8$ N. If the fuel capsule has surface $s = 5$ mm$^2$ = $5 \times 10^{-6}$ m$^2$ the capsule cover pressure is $p = F/s = 2 \times 10^{13}$ N/m$^2$ = $2 \times 10^8$ atm. This is gas dynamic pressure in the fuel capsule. The fuel compresses the shockwave from very high reactive pressure (see Fig. 8 - 10).

*Energy from ICF*. ICF can produce the energy

$$E = 1.6 \times 10^{-19} \frac{E_R M}{\mu m_p}, \quad [J],$$

where $E_R$ is the energy of single atom fusion, for D+T it is $E_R = 17.5$ MeV; $\mu = m_i/m_p$, for D+T $\mu = 2+3=5$. The fuel capsule having $M = 10$ $\mu g = 10^{-5}$ kg of a mixture D+T produces $3.34 \times 10^9$ J if all atoms take part in reaction. That is energy 84 liters of gas (benzene).

## Discussion

The suggested AB thermonuclear reactor has Lawson criterion in some order more then conventional current (2005) thermonuclear reactors (ICF). That strong increases either of three multipliers in Lawson criterion. That increases density $n$ up to two-three orders (see Figs.8-10). It increases the temperature $T$ in many times (it returns Bremsstrahlung radiation back to fuel pellet. This emission is main loss in current reactors), increasing the time of reaction $\tau$. For example, let us to estimate an increasing $\tau$ having laser impulse time $\tau = 10^{-9}$ s (it is time of laser impulse of conventional ICF), having a radius of combustion chamber 1 m and reflectivity coefficient $q = 0.98$. The reflective beam runs distance approximately $2 \times 50 \times 1 = 100$ m and spends the time $\tau = 100/(3 \times 10^8) = 333 \times 10^{-9}$ s. That means the pressure and time will be more by 333 times.

This show the suggested AB thermonuclear reactor may be a revolutionary jump in energy industry. That can decrease the power of requested laser, installation cost, and reactor size in tens time. The current inertial reactor needs only in installation of the prism-plasma reflector inside combustion chamber. The size of prism must be in some times less then size of the fuel capsule.

**Note**: In conventional ICF the initial (internal) radiation does not compress the plasma. When radiation came out of source (fuel pellet) and reflected or adsorbed by chamber surface that does not press on pellet surface. That is way the conventional inertial thermonuclear reactor has only losses from radiation. The offered AB reactor has the big benefits from radiation. The more radiation, the more benefits.

The efficiency from innovation for the magnetic thermonuclear reactor (MCF) is smaller because magnetic reactor works a small fuel density. For success, its work must change the place of fuel. Now the rare fuel gas (D+T) fills all volume of chamber. It must be located into small strong thin plastic tube under big pressure (see Fig. 5). In this case the fuel density can reach $n = 10^{26} \div 10^{27}$ 1/m$^3$ (or frozen fuel may be inside conductive matter, $n = 10^{28} \div 10^{29}$ 1/m$^3$). If the plasma reflectivity is high ($q > 0,99$), that is enough for thermonuclear ignition (Fig. 10) and keeping plasma under the radiation pressure and magnetic pressure. For current MCF the magnetic pressure is about $10^7$ N/m$^2$ (Eq. (9) for $B = 5$ T). For AB reactor the radiation pressure is about $10^{10} \div 10^{13}$ N/m$^2$ (millions atm). We can neglect the magnetic force and we may be design MCF reactor without very complex and expensive superconductivity magnetic system.

Some other author's ideas the reader find in [4]-[24].

## Acknowledgement

The author wishes to acknowledge R.B. Cathcart for correcting the author's English and other useful advice.